\documentclass[conference]{IEEEtran}

\usepackage{cite}
\usepackage{amssymb,amsfonts}
\usepackage{algorithmic}
\usepackage{graphicx}
\usepackage{textcomp}
\usepackage{xcolor}

\def\BibTeX{{\rm B\kern-.05em{\sc i\kern-.025em b}\kern-.08em
    T\kern-.1667em\lower.7ex\hbox{E}\kern-.125emX}}

  \usepackage{threeparttable}
  \usepackage{multirow}
  \usepackage{color}
  \usepackage{threeparttable}
  \usepackage{booktabs}
  \usepackage{cite}
  \usepackage{tabularx}
  \newcolumntype{Y}{>{\centering\arraybackslash}X}
  \usepackage{url}
  \usepackage{amsfonts}
  \usepackage{amssymb}
  \usepackage{algorithm}
  \usepackage{algorithmic}
  \usepackage[cmex10]{amsmath}
  \usepackage{color,soul}
  \usepackage{array}
  \newcommand{\etal}{{et al.}\xspace}
  \newcommand{\ie}{{i.e.}\xspace}

  \usepackage{url}
  \usepackage{xspace}

  \DeclareMathOperator{\RR}{\mathbb{R}}

  \newcommand{\fig}[1]{Fig.\,\ref{#1}}
  \newcommand{\tagapproach}{\textit{Tag}\xspace}
  \newcommand{\sensor}{\textit{ID-Sensor}\xspace}

  \usepackage{gensymb}

\usepackage{threeparttable}
\usepackage{multirow}
\usepackage{color}

\usepackage{booktabs}
\usepackage{tabularx}
\usepackage{algorithm}
\usepackage{algorithmic}

\usepackage[cmex10]{amsmath}

\usepackage{array}

\usepackage{url}
\usepackage{xspace}
\usepackage{epstopdf}

\usepackage{gensymb}

\ifCLASSOPTIONcompsoc
  \usepackage[caption=false,font=normalsize,labelfont=sf,textfont=sf]{subfig}
  \else
  \usepackage[caption=false,font=footnotesize]{subfig}
 \fi

\begin{document}

\title{Super Low Resolution RF Powered Accelerometers for Alerting on Hospitalized Patient Bed Exits}

\author{\IEEEauthorblockN{Michael Chesser\IEEEauthorrefmark{1},
Asangi Jayatilaka\IEEEauthorrefmark{1},
Renuka Visvanathan\IEEEauthorrefmark{2},
Christophe Fumeaux\IEEEauthorrefmark{3}
Alanson Sample\IEEEauthorrefmark{4} and\\
Damith C. Ranasinghe\IEEEauthorrefmark{1}
}
\IEEEauthorblockA{\textit{\IEEEauthorrefmark{1}Auto-ID Lab, School of Computer Science,
The University of Adelaide}, SA 5005 Australia\\
\{michael.chesser, asangi.jayatilaka, damith.ranasinghe\}@adelaide.edu.au }
\IEEEauthorblockA{\textit{\IEEEauthorrefmark{2}The Queen Elizabeth Hospital},
Woodville SA  5011, Australia\\
renuka.visvanathan@sa.gov.au}
\IEEEauthorblockA{\textit{\IEEEauthorrefmark{3}School of Electrical and Electronic Engineering,
The University of Adelaide}, SA 5005, Australia\\
christophe.fumeaux@adelaide.edu.au}
\IEEEauthorblockA{\textit{\IEEEauthorrefmark{4}School of Electrical Engineering and Computer Science,
University of Michigan}, MI 48109, USA\\
apsample@umich.edu}}

\maketitle

\begin{abstract}
Falls have serious consequences and are prevalent in acute hospitals and nursing homes caring for older people. Most falls occur in bedrooms and near the bed. Technological interventions to mitigate the risk of falling aim to automatically monitor bed-exit events and subsequently alert healthcare personnel to provide timely supervisions. We observe that frequency-domain information related to patient activities exist predominantly in very low frequencies.
Therefore, we recognise the potential to employ a low resolution acceleration sensing modality in contrast to powering and sensing with a conventional MEMS (Micro Electro Mechanical System) accelerometer.
Consequently, we investigate a batteryless sensing modality with low cost wirelessly powered Radio Frequency Identification (RFID) technology with the potential for convenient integration into \textit{clothing}, such as hospital gowns. We design and build a \textit{passive} accelerometer-based RFID sensor embodiment---\sensor---for our study. The sensor design allows deriving ultra low resolution acceleration data  from the rate of change of unique RFID tag identifiers in accordance  with the movement of a patient's upper body. We investigate two convolutional neural network architectures for learning from raw \textit{RFID-only} data streams and compare performance with a traditional shallow classifier with engineered features. We evaluate performance with 23 hospitalized older patients. We demonstrate, for the first time and to the best of knowledge, that: i) the low resolution acceleration data embedded in the RF powered \sensor data stream can provide a practicable method for activity recognition; and ii) highly discriminative features can be efficiently learned from the raw RFID-\text{only} data stream using a fully convolutional network architecture.
\end{abstract}


\section{Introduction}\label{sec:introduction}
Approximately 30-50\% of older people living in long-term care institutions fall each year \cite{who_falls_site}.  The world population is rapidly ageing and in the year 2050 approximately 2 billion people will be 60 years of age or older \cite{organization_ghana_20142}. Therefore, we can expect ageing associated issues such as increased risk of falling to become more prevalent.
Falls lead to many adverse consequences for the patients apart from physical injuries such as anxiety, depression and loss of independence~\cite{rapp_epidemiology_2012, robinovitch2013video}.
Furthermore, falls are costly because they increase the length of hospital stays. A recent study estimated the total medical costs for falls in 2015 at approximately \$50 billion USD in the United States alone~\cite{florence2018medical}.
In residential care and hospitals, falls commonly occur near and around the patients' beds \cite{rapp_epidemiology_2012}.
A recent study, a first of its kind, conducted an in-depth analysis of video surveillance recordings over three years and revealed that people fall as a result of getting out of bed and also as a result of walking~\cite{robinovitch2013video}.

Automatically recognizing patients leaving their bed in real time provides an opportunity to intervene and supervise unattended patients. A common strategy to  provide targeted care to older patients in hospitals is by using automatic alarm systems.  Alarm systems issue a warning when a patient is getting out of bed with the aim of staff promptly attending to the patient and thereby potentially reducing the risk of a fall or rendering immediate assistance in case of a fall~\cite{sahota2013refine,shorr2012effects,visvanathan2017effectiveness}. Such systems, focused on prevention are more desirable than systems focused on detecting a fall after the event has \textit{already} occurred~\cite{visvanathan2017effectiveness}. Current approaches to recognize bed-exits using pressure sensors---for example, pressure mats~\cite{bruyneel_detection_2011, arcelus2011measurements}---are shown to be ineffective in reducing falls in clinical trials~\cite{capezuti_bed-exit_2009}.  Older people have expressed privacy concerns with the use of camera based technologies~\cite{demiris2008senior}. Recent studies have explored  battery-powered body worn sensors for patient activity recognition~\cite{wolf_development_2013,visvanathan2017effectiveness}; however, these sensors are expensive\footnote{MbientLab Bluetooth sensor pricing (these prices \textit{do not} include the cost of coin cell batteries) \url{https://mbientlab.com/pricing/}}, require manual attachments, maintenance and battery replacements, and strapping to the body as in~\cite{wolf_development_2013}.

\begin{figure}
	\centering
		\subfloat[]{
			\centering
			\includegraphics[width=0.45\linewidth]{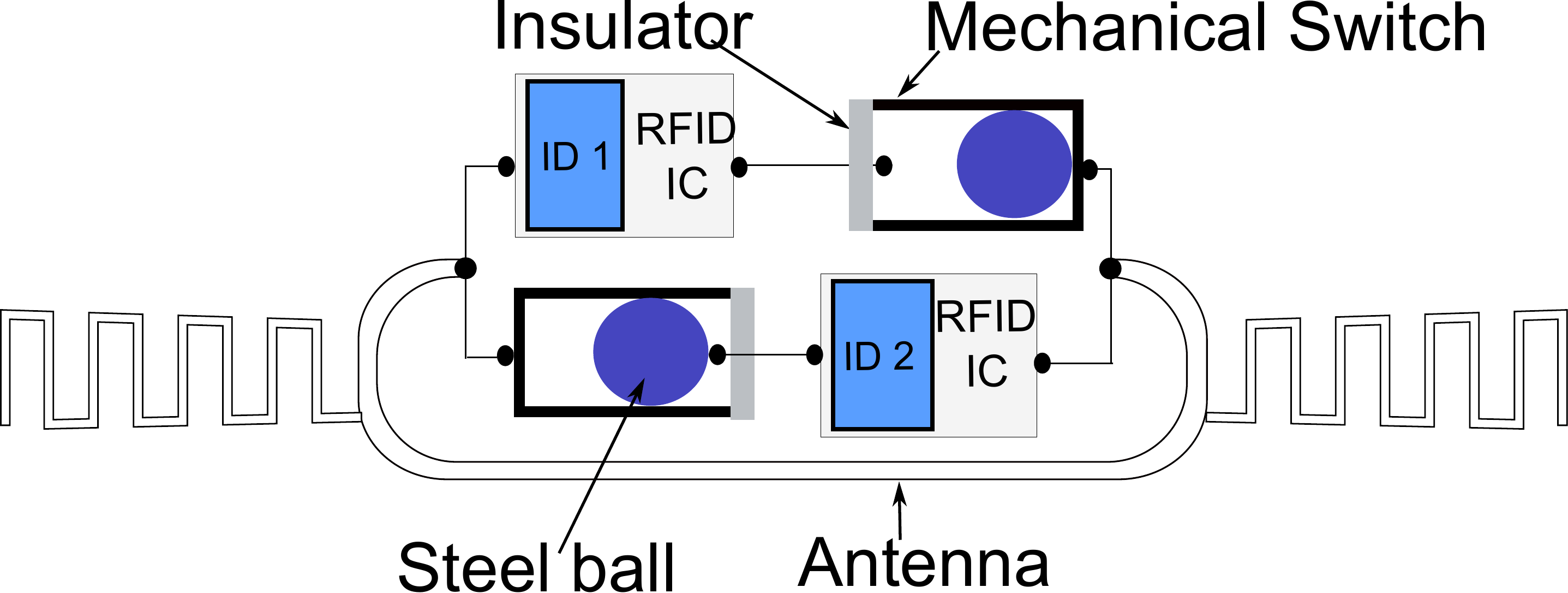}
			\label{fig:logical}
		}
		\subfloat[]{
			\includegraphics[width=0.45\linewidth]{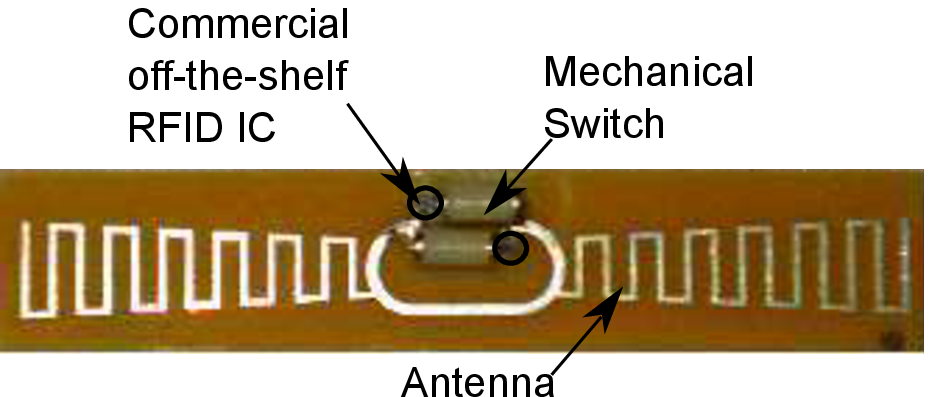}
			\label{fig:alpha}
		}

		\subfloat[]{
			\includegraphics[width=0.25\linewidth]{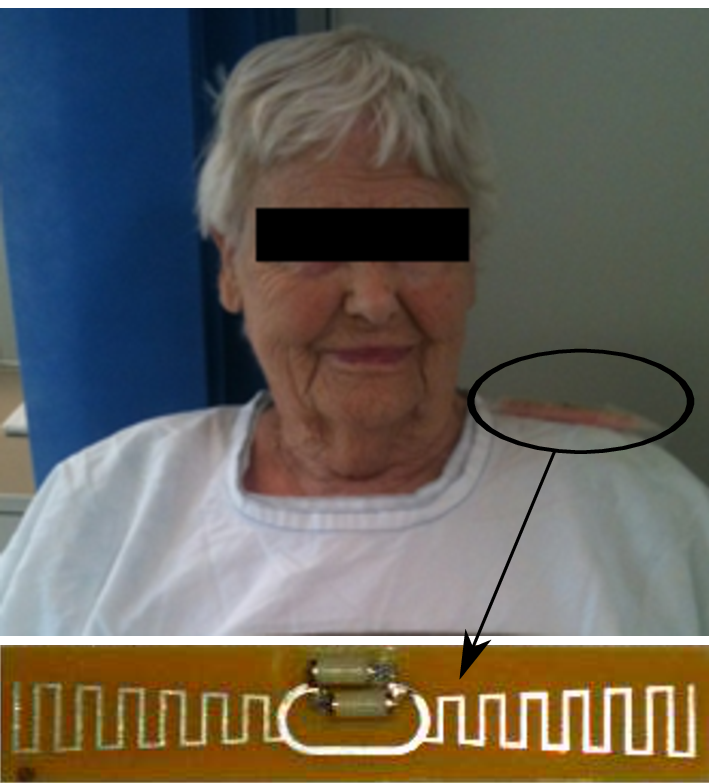}
			\label{fig:patient}
		}
		\subfloat[]{
			\includegraphics[width=0.73\linewidth]{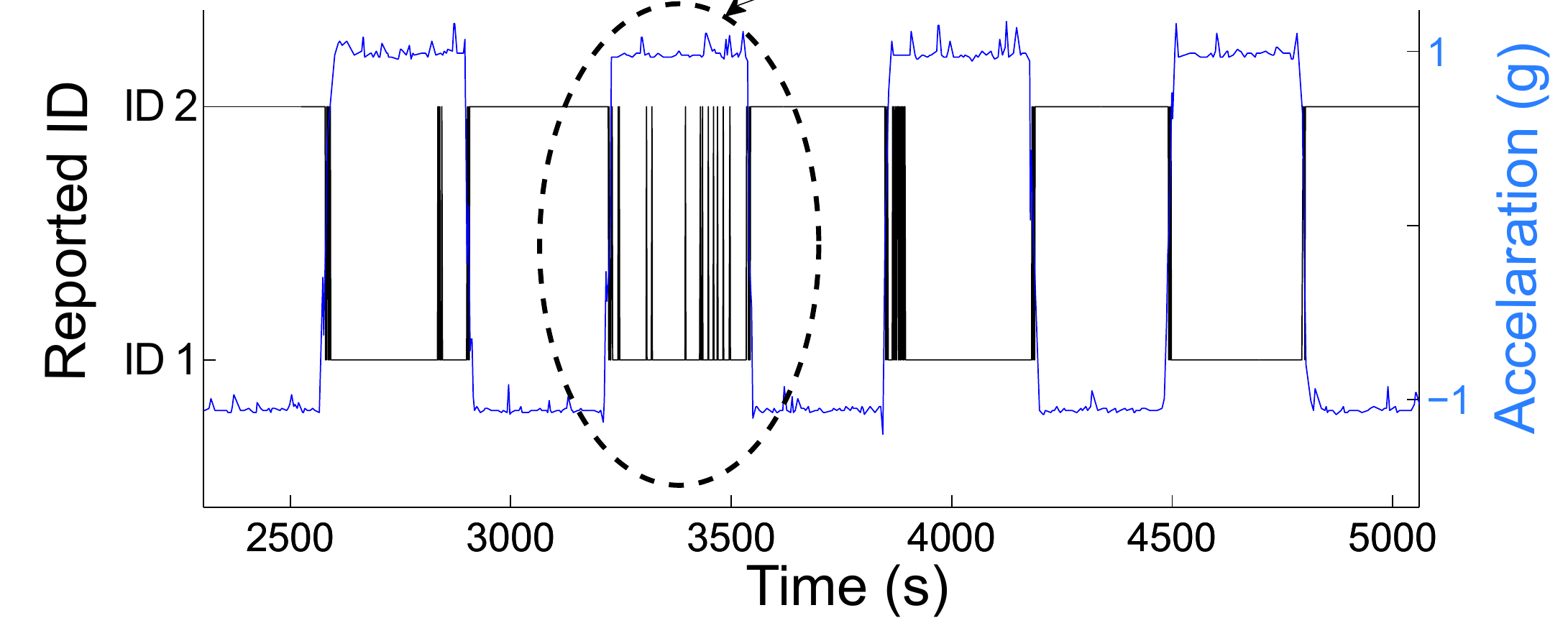}
			\label{fig:accelcomparison}
		}

		\caption{ (a)\, Design of the RFID sensor tag.  Each RFID Integrated Circuit (IC) has a unique tag ID. As a result of only one RFID IC being connected to the antenna,  only one ID will be reported by the RFID platform at a given time. This figure illustrates the state of the   \sensor when ID 2 is reported.   (b)~\sensor: The batteryless RFID sensor tag (\sensor) prtotype used in our experiments. (c)\,A patient wearing the batteryless RFID tag attached at the shoulder level to a hospital garment. (d)\, Comparison of the ID reported by the \sensor against acceleration values obtained from a MEMS accelerometer. }
\end{figure}

In contrast, we postulate an alternative. We observe that patient activity related information in the frequency-domain is found in very low frequencies of typically less than 4~Hz~\cite{najafi2003ambulatory}. Therefore, we recognise the  potential to use a low resolution acceleration sensor to gather human motion information in contrast to powering and sensing with a conventional MEMS (Micro-Electro-Mechanical Systems) accelerometer. Consequently, we design an RF (radio frequency) powered or passive sensing modality with low cost batteryless Radio Frequency Identification (RFID) technology with the potential for convenient integration into \textit{clothing}, such as hospital gowns. In particular, we investigate the efficacy of embedded acceleration data extracted from the modulation of two unique identifiers in an RFID data stream. The change in identifiers or IDs in accordance with human motion data is derived from a tag with two commercial off-the-shelf (COTS) RFID circuit modules as illustrated in Figure~\ref{fig:alpha}.

There are many advantages to using passive wearable UHF RFID for capturing the movements of older patients: \textbf{\textit{i)}}~passive devices have the potential for an indefinite operational life without requiring maintenance or battery replacements; \textbf{\textit{ii)}}~wearable RFID technology addresses the problem of distinguishing individual patients from multiple others, faced by most device-free sensing schemes and allows individualizing bed-exit alarms to match patient needs over time;  and \textbf{\textit{iii)}}~RFID tags are low-cost (7-15 U.S. cents)~\cite{passiveRFIDJournal}, hence, disposable to support infection control protocols in hospitals. Furthermore, machine washable RFID tags are now commercially available and can be easily woven into hospital garments~\cite{fujitsu-laundry-tag}; thus, creating possibilities for unobtrusive monitoring of patient activities. Unobtrusiveness has been identified as a key acceptance criteria by older people~\cite{govercin2010defining,roberto-pone.0185670}; a necessary condition for translating technology into practice. Most notably, we see an increasing technology trend to integrate RFID technology with textiles~\cite{kiourti2018rfid}.

We have seen demonstrated capability to use commercially available passive UHF RFID technology for recognizing human interactions with RFID tagged objects \cite{jiang2006unobtrusive, li2015idsense,jayatilaka2017real}, tracking of objects or people~\cite{adame2018cuidats,yao2018efficient}, as well as use of large number of body-attached RFID tags for human motion tracking~\cite{wang2018rf, jin2018towards}. However, our study is a \textit{first-of-its-kind}. To the best of our knowledge: \textit{\textbf{i}}) the potential to extract and use low resolution acceleration data in the form of ID modulations from an extremely low power method for human activity recognition; and \textit{\textbf{ii}}) a practicable means for recognizing bed egress motion for a clinical application with a worn RFID tag and the evaluation with a target demographic of frail older hospitalized people, have not been previously investigated.

\subsection{Contributions}
Our main contributions of this paper are given below.
\begin{itemize}
	\item We investigate a new and pragmatic human motion sensing approach. We  observe  that  patient activity related information in the frequency-domain exist in very low frequencies. We construct a prototype RFID device---\sensor---using only low cost commercially available RFID Integrated Circuits (ICs) to capture ultra low resolution acceleration data. We show that the \sensor provides a unique ability to capture human motion information without a conventional accelerometer.

	\item We share the finding that deep convolutional neural network (CNN) architectures can automatically learn discriminative features from our unique \sensor data with embedded low resolution acceleration information in an RFID-\textit{only} data stream. To the best of our knowledge, only one other study has considered the problem of human activity recognition from \textit{raw} RFID-only data streams using a deep learning paradigm~\cite{LiDNN-RFID2016}.

	\item We demonstrate, for the first time and to the best of our knowledge, the capability to employ a low cost body-worn passive UHF RFID tag for sensing of hospitalised patient activities. Our pilot study is performed with 23 hospitalised older patients. In particular, our study is conducted in a realistic experimental setting and with a demographic of participants intended for the application.

	\item We release a new activity dataset collected from hospitalized older people to the research community---see~\cite{githublink2018}.
\end{itemize}

We defer related work to Section~\ref{sec:relatedwork}. In the following sections we describe our sensor construction and operation, followed by the experimental design and the machine learning approaches we formulate for the activity recognition task. We discuss our results and conclude our paper in Section~\ref{sec:discuss-conclude}.

\section{Passive RFID-based ID-Sensor}\label{sec:technology}

Passive RFID tags harvest energy radiated by an RFID reader antenna and once successfully powered, responds by backscattering the Radio Frequency (RF) signals back to the RFID reader antenna.
Apart from the unique electronic identifier sent from a tag, modern RFID readers are able to measure detailed RF communication-related properties such as received signal strength expressed as RSSI and phase difference of the received signal. While information extracted from RSSI has been exploited in the past, the exploitation of information related to \textit{changes in phase} information are rarely explored for human activity recognition. We aim to exploit both of these information sources. Notably, both phase and RSSI related information is extracted at no additional computational or power burden to the tag and thus does not impact the reading range or the performance of the tag.
\vspace{1mm}

\noindent\textit{Our goal is to create and exploit another such information source for human activity recognition}.
\vspace{1mm}

We employ low-cost batteryless RFID technology to build our \sensor to capture low frequency information related to human motion. We developed this prototype based on~\cite{philipose_battery-free_2005}. The platform consists of two antiparallel mechanical tilt switches and two RFID Integrated Circuits (ICs) attached to a single Radio Frequency (RF) antenna on a Printed Circuit Board (PCB) substrate (see \fig{fig:logical} and \fig{fig:alpha}). Since the mechanical switches are attached opposite to each other, only one RFID IC is attached to the antenna at a time. Each RFID IC has a unique tag ID (\textsf{ID1} or \textsf{ID2}). Starting from the case where the first RFID IC is connected, flipping (or shaking) the tag will result in the first IC becoming disconnected, and the second RFID IC becoming connected to the antenna. At a given time, the ID stored in the RFID IC connected to the antenna will be reported by the platform. Thus, tilt and accelerometer data are encoded as  \textit{modulations} between the two IDs reported back to the reader.

The key concept here is the exploitation of two tiny batteryless UHF RFID tag ICs in a configuration that allows switching between the ICs based on the direction of the gravitational force vector in relation to a human body reference frame.
Most notably, the accelerometer in essence consumes no electrical power, or more accurately, is limited to the negligible Ohmic losses of the switching element. Importantly, the \sensor uses low-cost readily available components, is mass producible, and has RF performance equivalent to conventional single IC RFID tags. Furthermore, patients are precisely and automatically identified by the unique electronic identifiers stored in the worn RFID tags.

\begin{figure}[t]
	\centering
	\centering
	\subfloat[]{
		\centering
		\includegraphics[trim={0 0 0 0},clip,width=0.37\linewidth]{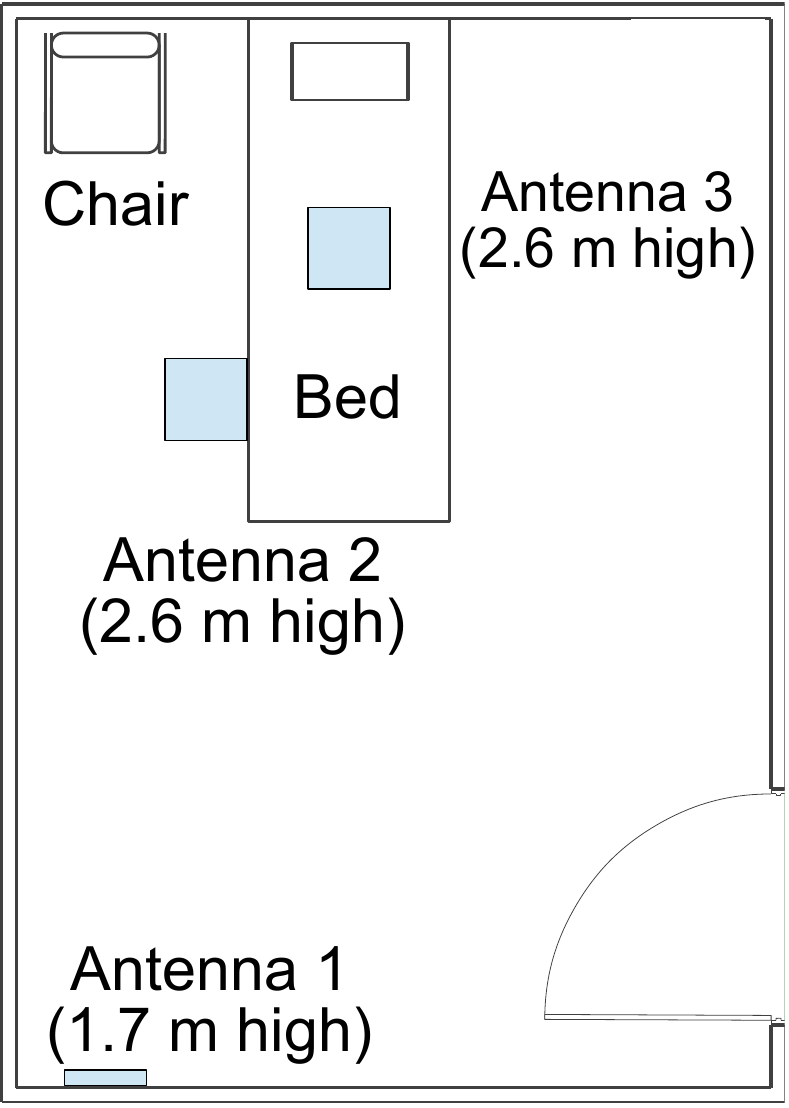}
		\label{fig:room}}
	\subfloat[]{
		\centering
	\includegraphics[trim={0 0 0 2cm},clip,width=0.37\linewidth]{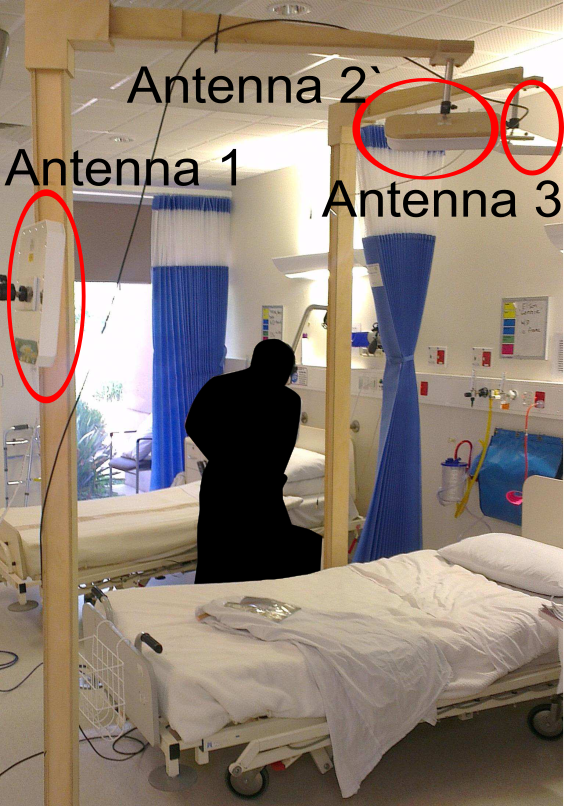}
		\label{fig:room_actual}}
	\caption{  (a)\, Typical layout of a hospital bed room. The picture shows a double-bed room used in our experimental study and indicates the reader antenna locations. (b) Experimental set up in a double bed room. }
\end{figure}

Figure~\ref{fig:accelcomparison} compares the low resolution acceleration information received from the \sensor with acceleration values obtained using a MEMS accelerometer (ADXL330). This was obtained by rotating both \sensor and the accelerometer at an identical rotational velocity.  We can see that the ID reported by the \sensor is capable of representing the direction of the gravitational force vector and rate of rotation or angular acceleration. However, it is also evident that the ID reported by the \sensor can sometimes result in noisy measurements as highlighted in \fig{fig:accelcomparison}. This is due to the steel ball failing to disconnect at times and as a result connecting both RFID ICs to the antenna simultaneously.
\vspace{1mm}

\noindent\textit{Therefore, our study will investigate the ability to use this noisy low resolution acceleration information for human activity recognition problems.}
\vspace{1mm}

The consideration of whether the noisy acceleration data provides any additional information forms the basis for the two approaches we will investigate:

\begin{itemize}
	\item \textbf{\sensor Approach:} Here, we utilize the embedded acceleration data in the form of changes in IDs from our \sensor in addition to RSSI and phase information.
	\item \textbf{\tagapproach Approach}: We ignore the ID modulation information and we treat \sensor as a simple commercial off-the-shelf (COTS) batteryless UHF RFID tag. Hence we do not distinguish between \textsf{ID1} and \textsf{ID2} and instead treat them as a single identifier.
\end{itemize}

\section{Experimental Study}\label{sec:study-participants-and-data-collection}
The participants for our experimental study were inpatients at the Geriatric Evaluation and Management Unit of the Queen Elizabeth Hospital, South Australia. The patients selected for the study were able to consent to the study and mobilize independently. This study had ethics approval from the human research ethics committee of the Queen Elizabeth Hospital, South Australia (2011129). We describe the details of the data collection and experimental settings below.
\vspace{1mm}

\noindent\textbf{Participants: } Twenty three older participants were recruited (age: 84.4 $\pm$ 5.3 years, height: 1.68 $\pm$ 0.09 m) for the study with the help of geriatricians. All participants provided informed consent and no honorarium was paid.
The study was completed over a six-month period where each trial with each volunteer lasted between 60 to 90 minutes. The \sensor was attached to the loosely fitted hospital gown and over a participant's shoulder as shown in \fig{fig:patient}.
\vspace{1mm}

\noindent\textbf{Settings: } The data collection occurred in the individual rooms of patients consenting to the study. These patient rooms included both double and single bed rooms. The furniture and, hence, the antenna deployment in all the patient rooms were similar. The generic deployment of antennas used in the experiments is illustrated in \fig{fig:room} and \fig{fig:room_actual}. During trials the position of the back rest on a bed was not fixed and was generally elevated slightly to suit the personal comfort of individual patients. Although the room setting was mostly fixed during the experiment, movement of other people (such as nurses) was not restricted.

We used an Impinj Speedway Revolution reader operating at the regulated Australian RF frequency band of 920-926 MHz and  a maximum regulated power of  1~W. The communication between the RFID tag  and the reader is governed by the  ISO-18000-6C air interface protocol.
In this study, three antennas  were attached to the RFID reader and strategically deployed in  patient's rooms (see \fig{fig:room} and \fig{fig:room_actual}).  The antenna deployment was designed to illuminate the area covering the bed and chair.  The read distance of UHF RFID tags in free space is generally 10\,m.  In  the  experimental set-up, the read rate (or sampling rate) of the \sensor  was approximately 20 reads per second.
\vspace{1mm}

\noindent\textbf{Data collection:} Since our study participants were hospitalized older frail patients, it was observed that most of their time was spent lying in bed. Therefore, in order obtain sufficient amount of information for  bed-exit events, and to minimize the physical and mental stress for the participants, the study was conducted using broadly scripted activity routines to allow us to obtain an adequate number of bed exit events.

The participants were instructed, prior to each trial, to lie on the bed in a manner most natural and comfortable to them.  They were requested to get out of the bed during the experiment and no specific instructions were given about how or when to get out of the bed.  As they were frail patients, the number of times each patient got out of the bed was not fixed and was dependent on their physical abilities. They were allowed to perform other  activities while  \textit{in-bed} as well as while \textit{out-of-bed}.  Typical \textit{in-bed} activities involved lying, sitting, watching a television, drinking and reading. Typical \textit{out-of-bed} activities involved standing and walking. Given the very limited number of activities a patient can perform in a hospitalized environment, we believe these activities represent adequate class diversity. Given the varying physical abilities of our patient group, we also benefited from very high intra-class diversity.
A researcher annotated  the activities  being undertaken. Given our interest in bed egress motions, two ground truth activity labels were considered for our study: i)~\textit{in-bed}; and ii)~\textit{out-of-bed}.
\vspace{1mm}

\noindent\textbf{Data set:} The total data set contains 234,764 tag readings which  included 109,141 \textit{in-bed} and  125,623 \textit{out-of-bed} related tag readings. Read rate (or sampling rate) was generally between 5-20 reads per second. The data set included 70 bed-exit events (\ie transitions from \textit{in-bed} to \textit{out-of-bed}).

\begin{figure}[t]
\centering
	\includegraphics[width=0.8 \linewidth]{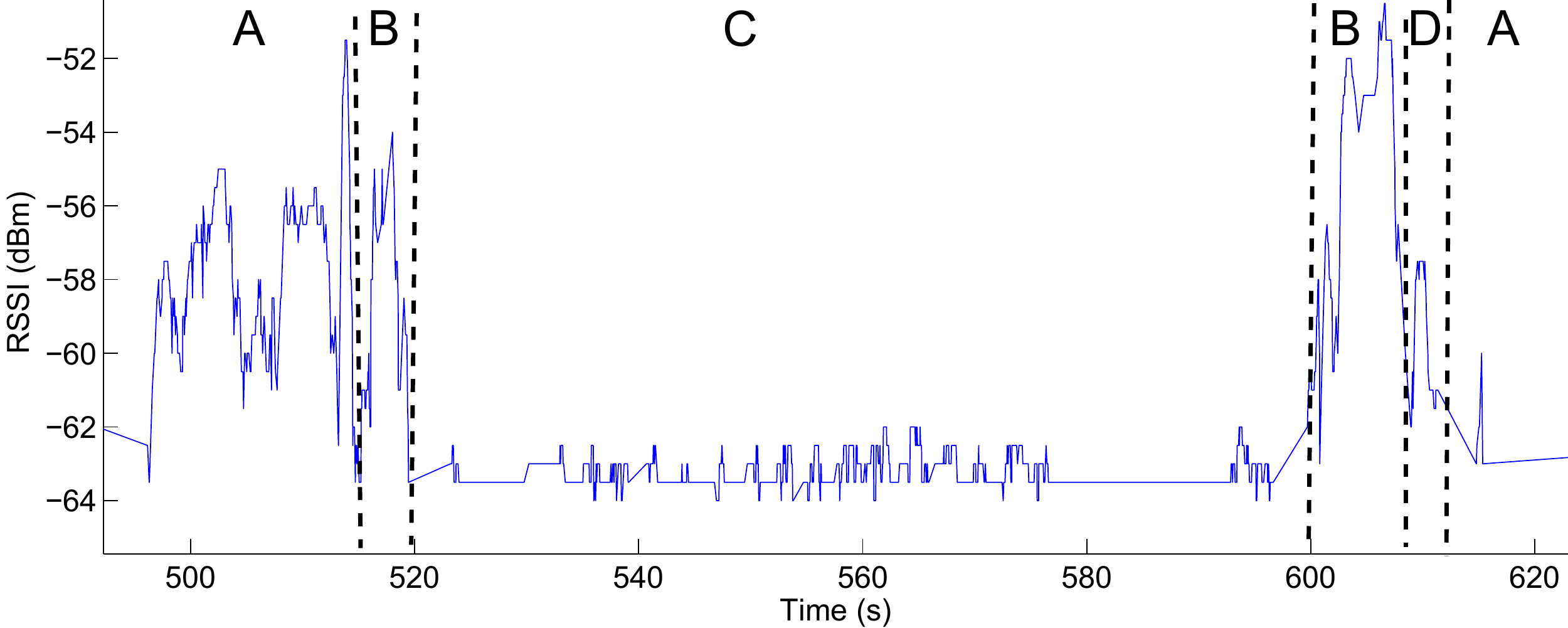}
	\caption{Typical variations in RSSI values for \textit{in-bed} and \textit{out-of-bed} activities. Here, A: \textit{walking}, B: \textit{Sitting on bed}, C:\textit{ lying on bed} and D: \textit{Standing}}
	\label{fig:minmax}
\end{figure}

\begin{figure}[t]
	\centering

	\includegraphics[width=0.8\linewidth]{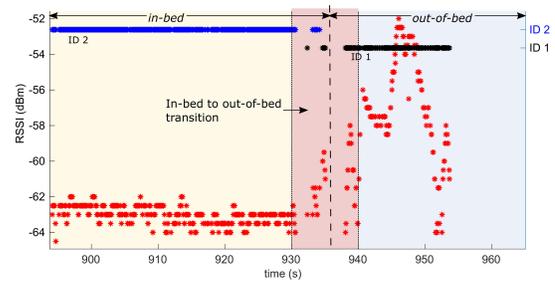}
		\caption{
		Capture of a bed-exit posture transition of a patient where the ground truth changes from \textit{in-bed} to \textit{out-of-bed}. Changes in ID information embed low resolution accelerations using the \sensor. We can also observe typical RSSI patterns associated with the activated ID of the \sensor prior to, during and after the activity transition.}
		\label{fig:state1}
\end{figure}

\section{Approach}
A bed-exit event is recognized as a transition from \textit{in-bed} to \textit{out-of-bed}. Once a bed-exit is identified, interested parties, such as caregivers, can be notified, by means of a pager message or phone notification as discussed in \cite{roberto-pone.0185670,visvanathan2017effectiveness}. Our approach to recognize bed-exits is built around the observations that: i) human motion information can be extracted from the channel state measurements, RSSI and phase, made by an RFID reader; and ii) low frequency information associated with patient movements from the \sensor where the data stream from the reader can be also be used to extract low resolution acceleration information. Importantly, we can extract all three information sources without a power burden on the tag and through standard interrogation of the \sensor. However, all of these measurements are noisy; the challenge is to learn the hidden patterns to recognize bed egress motions from noisy data. We discuss these information sources briefly.
\vspace{1mm}

\noindent\textbf{Signal strength information:} RSSI is an indicator of the  power of the tag signal received by an RFID reader antenna~\cite{nikitin2010phase}. RSSI is predominantly affected by the distance between the reader and the tag as well as the orientation of the tag antenna.
Based on the Friis transmission equation, RSSI of a backscatterd signal that is captured by an RFID reader has the form of $P_{t}G^{2}_{t}G^{2}_{path}K$.

Here, $P_t$ is the output power of the reader, $G_t$ is the gain of the reader antenna, $K$ is the  backscatter gain. The $G_{path}$ is the one-way path gain of the deterministic multipath channel determined as  $G_{path} \ =\ \left( \frac{\lambda}{4 \pi R} \right)^{2} \mid H\mid $.

Here, $R$ is the line of sight distance between the tag and the reader antenna, and $H$ is the channel response due to multipath and channel absorption characteristics.
For instance $H \propto e^{-\alpha R}$ where $\alpha$ is the absorption coefficient of the medium.

We can see that although RSSI is sensitive to channel characteristics, it is predominately determined by $R$ as $RSSI \propto 1/R^4$.

Figure~\ref{fig:minmax} illustrates data collected using reader antenna 3 for a sequence of activities related to getting into bed and then getting out of bed.
From \fig{fig:minmax} we can observe that RSSI is sensitive to movements. For instance, the posture transition from \textit{lying on bed}  to  \textit{sitting on bed} (\ie in  \fig{fig:minmax}, from C to B) has resulted in a notable increase in RSSI value while the posture transition from \textit{sitting on bed} to \textit{standing} (\ie in ~\fig{fig:minmax}, from B to D during 600\,s to 620\,s) has resulted in a decrease in the RSSI value.
\vspace{1mm}

\noindent\textbf{\sensor information: } We have illustrated in  \fig{fig:accelcomparison} and discussed in Section~\ref{sec:technology} the capability of obtaining acceleration data by considering the ID modulations within the RFID data stream. As observed in \fig{fig:state1}, changes in the reported IDs are likely to be observed for a patients upper body movement such as during a bed egress activity. This is a consequence of the sensor moving in accordance with a patient's upper body movements and rotations that changes the orientation of the sensor's mechanical switches with respect to the gravitational force vector resulting in modulations between \textsf{ID1} and \textsf{ID2} of the sensor. These modulations capture low frequency acceleration information associated with human motions. Further, we observe that RSSI patterns captured by each activated ID generally differ according to a patient's lying and standing posture as well as during a posture transition (see \fig{fig:state1}). Therefore, RSSI patterns specific to each ID provide further movement information.
\vspace{1mm}

\begin{figure}[t]
	\centering
	\includegraphics[trim={0 0.5cm 0 0},clip,width=\linewidth]{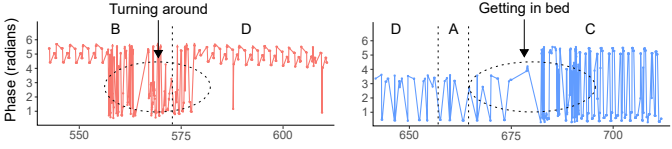}
	\caption{An extract of variations in instantaneous phase values for \textit{in-bed} and \textit{out-of-bed} activities of two antennas (red: antenna 1, blue: antenna 3). Where A: \textit{walking}, B: \textit{Sitting on bed}, C:\textit{ lying on bed} and D: \textit{Standing}}
    \label{fig:phase_plot}
\end{figure}

\noindent\textbf{Phase information:} Phase estimates from an RFID reader is a measure of the phase angle between the RF carrier transmitted by the reader and the return signal from the tag. RFID readers performs frequency hopping from one channel to another. As a result, phase values are dependent on the carrier frequency.  Phase estimates are related to two motion information sources: \textit{\textbf{i}}, for a given frequency, the phase of the tag signal at two different time instances can estimate the tag's radial velocity as $ V_r \propto \frac{\partial  	\Psi}{\partial t}$; and \textit{\textbf{ii}}) the distance between an RFID reader antenna and the RFID tag is proportional to the partial derivative of the phase with respect to the derivative of channel frequency as   $d \propto \frac{\partial  	\Psi}{\partial f}$. Given the sensitivity of phase information to small changes in distance, we can expect phase data to capture activity transition information. Figure~\ref{fig:phase_plot} shows an extract of variations in phase information across two activity transitions in our dataset. We can observe that RF phase ($\Psi$) is generally affected by even small movements of an RFID tag while RSSI is primarily affected by much larger movements.

\subsection{Classification Problem}
We are interested in determining the associated activity label (\textit{in-bed} or \textit{out-of-bed}) for each \sensor reading. We treat the problem of determining whether a patient is \textit{in-bed} or \textit{out-of-bed} to be a binary classification problem. Subsequently, a bed-exit event is recognized as a change in  classification from \textit{in-bed} to \textit{out-of-bed}.

The data sequence collected is a time series and a  single tag reading consists of the 4-tuple: i)\,reader antenna ID (aID); ii)\,RSSI; iii)\,Phase; and iv)\,tag ID (ID).  It is important to  transform the received data sequence into a suitable representation before applying activity-based machine-learning models. The most common strategy is to segment the time series. We used a fixed time sliding window, however,  the selection of segment size ($\delta s$) is an empirical process determined by each algorithm. Our goal is to investigate a classical machine learning method with a feature learning method based on state-of-the-art neural network architectures used with kinematic sensor data. We recognize that the nature of RFID data is significantly different from kinematic sensor data and feature learning from RFID data streams for activity recognition remains to be explored. Therefore, we use the classical machine learning method of Logistic Regression (LR) with engineered features as a benchmark to compare with two deep neural network architectures.

\subsection{Logistic Regression}\label{sec:features}
In this study, we used a classical machine learning algorithm capable of generating probabilistic models using feature vectors extracted from segments. LR assumes that the training data,
$(x_{t},y_{t})$ where ${t \in \mathbb{N}}$, $x_{t} \in \RR^d$   is the feature vector and $y_{t} \in \{ -1, +1 \}$, is the class label, are independent and identically distributed. LR models the conditional probability $Pr(y_{t}=1| x_{t})$ as follows:
\vspace{-1mm}
	\begin{displaymath}
	Pr(y_{t}=1| x_{t}) = \left(   \frac {e^{<w,x_{t}>}}{1+ e^{<w,x_{t}>}}   \right) =  \left(   \frac {1}{1+ e^{-(<w,x_{t}>)}}   \right)
	\end{displaymath}

\noindent here $w$ is the learned model.

We engineered a number of features using the information available in a given segment with respect to: \textit{\textbf{i}}) RSSI; and \textit{\textbf{ii}}) correlation between participant's movements and antennas capturing tag readings. Additional features utilizing one-bit acceleration were engineered for the \sensor approach (see Table~\ref{tbl:features}).

\noindent\textbf{Features based on RSSI: } For a given segment $S_i$ with respect to a given tag reading, \textit{i}, we used the features summarized in Table~\ref{tbl:features} and also used in~\cite{Wang2013,roberto-pone.0185670}. In order to further capture the changes in RSSI for short distance movements, we introduce new a binary feature $M \in {[0,1]}^{|\mathcal{A}|}$ to determine whether  a patient is moving towards or away from a fixed antenna within $S_i$  where $M_k$ is defined as:
\vspace{-4mm}
\begin{align}
M_{k}=\mathbf{1}_{[t_{max(S_i(RSSI^k))} > t_{min(S_i(RSSI^k))}]},
\end{align}

\noindent where $k \in \mathcal{A}$ and $\mathbf{1}_{x}$ assumes 1 if $x$ is true and 0 otherwise.  $\mathcal{A}$ is the set of antennas in the deployment.
Although the proposed binary feature $M_k$ can be used to identify human movements within a segment, it cannot be used to recognize trends in the RSSI values over a longer period. To capture these patterns we considered a longer segment $\mathfrak{S}_i$ of size $3\delta s$ with three equal sub-segments $\mathfrak{S}_i^j, j = 1,...,3$. We consider the mean value of RSSI in each sub-segment. Since $\mathfrak{S}_i^3 = S_i$ we only include mean RSSI value for $\mathfrak{S}_i^j, j = 1,2$.
\begin{table}
	\centering
	\caption{Features Extracted from a Segment}
	\label{tbl:features}
	\resizebox{1\linewidth}{!}{
		\begin{threeparttable}
			\begin{tabular}{p{0.36\linewidth} p{0.77\linewidth}}
				\hline
				Notation                                            & Description                                                        \\ \hline
				\multicolumn{2}{l}{\textbf{RSSI based Features}}                                                                               \\ \hline
				$RSSI_i$                                             & Most recent RSSI value                       \\
				mean ($S_i(RSSI^k)$)                     & Mean RSSI value                                                     \\
				max ($S_i(RSSI^k)$)                 & Maximum value of the RSSI                                           \\
				min ($S_i(RSSI^k)$)                  & Minimum value of the RSSI                                           \\
				std($S_i(RSSI^k)$)                       & Standard deviation of the RSSI                                      \\
				$M_k$                               & Whether the maximum RSSI value is followed by the minimum RSSI value \\ \hline
				\multicolumn{2}{l}{\textbf{Phase based Features}}                                                                               \\ \hline
				$\Psi_i$                                             & Most recent phase value                       \\
				median($S_i(CFPR^k)$)                               & Median of the CFPR \\
				sum($|S_i(CFPR^k)|$)                               & Sum of the absolute values of the CFPR \\
				std($|S_i(CFPR^k)|$)                               & Standard deviation of the CFPR \\
				    \hline
				\multicolumn{2}{l}{\textbf{Event based Features}}                                                                             \\ \hline
				$RC_k$                                   & Relative read event count per antenna                                     \\
				$\omega_k$                                            & Antenna which corresponds to the majority of events in $S_t$       \\
				$aID_i$                                             & The antenna ID corresponding to the most recent tag reading        \\ \hline
				\multicolumn{2}{l}{\textbf{Features Specific to \sensor approach}}                                                       \\ \hline
				$a_i$                                               & ID of the most recent tag reading                               \\
				$|S_i(a=x)|/|S_i|, x=1,2$                           & Relative ID count                                               \\ \hline
			\end{tabular}
			\begin{tablenotes}
	    		\item[*] $k=1,2,3$ (antenna number), $i$ is number of the respective tag reading
		    \end{tablenotes}
		\end{threeparttable}

	}
\end{table}
\vspace{1mm}

\noindent\textbf{Features based on correlation between participant's movements and antennas capturing the tag readings: }
It was observed that the antennas that collect readings differ depending on the location and movements of the patient. The RFID antenna facing the RFID tag are most likely to both power and collect data from the tag.

First, we considered a feature that indicates the relative count of read events for each antenna  ($RC \in \mathbb{R}^{|\mathcal{A}|}$)  within $S_i$ which is calculated as:
\vspace{-1mm}

\begin{align}
 RC_{k}=\frac{|S_i(aID=k)|}{|S_i|},
 \end{align}

 \noindent where $k \in \mathcal{A}$. Second, we introduced a vectorized binary feature, $\omega \in \{0,1\}^{|\mathcal{A}|}$, to identify the antenna which corresponds to the majority of events in $S_i$, which takes values as:
\begin{align}
\omega_{k} =
\begin{cases}
1 & \text{if } k= \underset{l \in A}{\arg\max}\,\,RC_l\\
0       & \text{otherwise }
\end{cases}
\end{align}
where $\omega_{k}$ represents the $k^{th}$ position of the vector $\omega$ which corresponds to the  antenna $k$.
\vspace{1mm}

\noindent\textbf{Phase-based Features: }In our study we used statistical features derived from phase measurements from each antenna as in~\cite{li2015idsense,jayatilaka2017real}. These features are summarized in Table~\ref{tbl:features}.

\begin{figure*}[t]
    \centering
	\subfloat[]{
		\centering
		\includegraphics[width=0.49\linewidth]{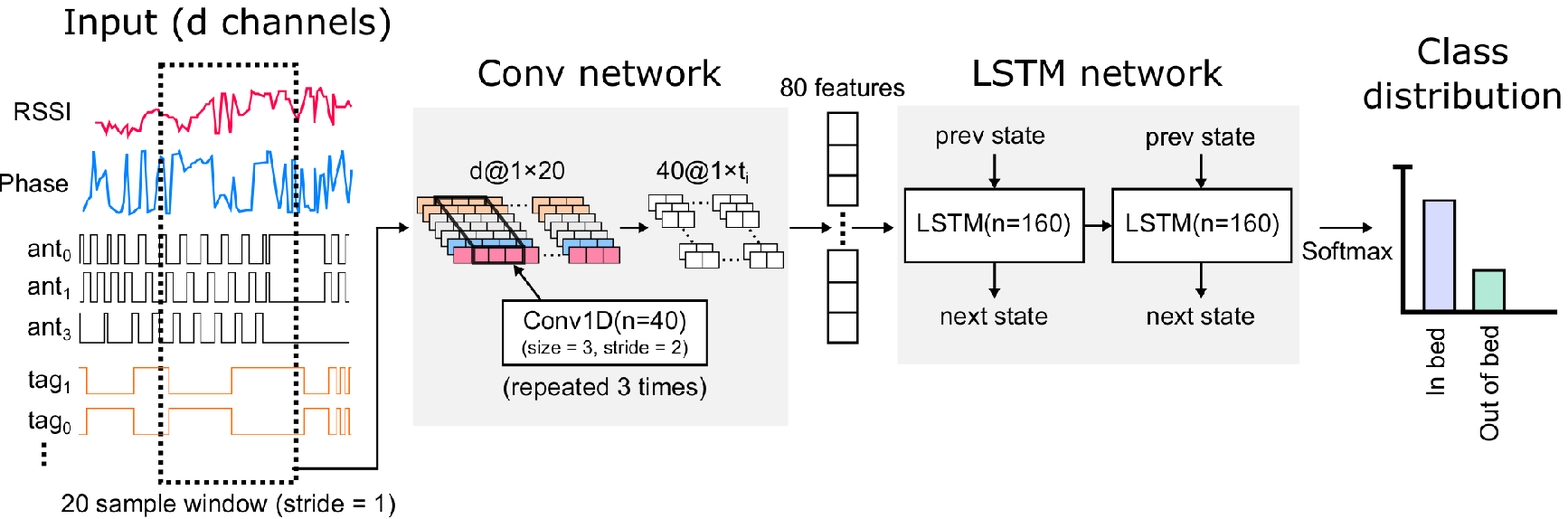}
		\label{fig:cnn_lstm_plot}

	}
	\subfloat[]{
		\centering
		\includegraphics[width=0.48\linewidth]{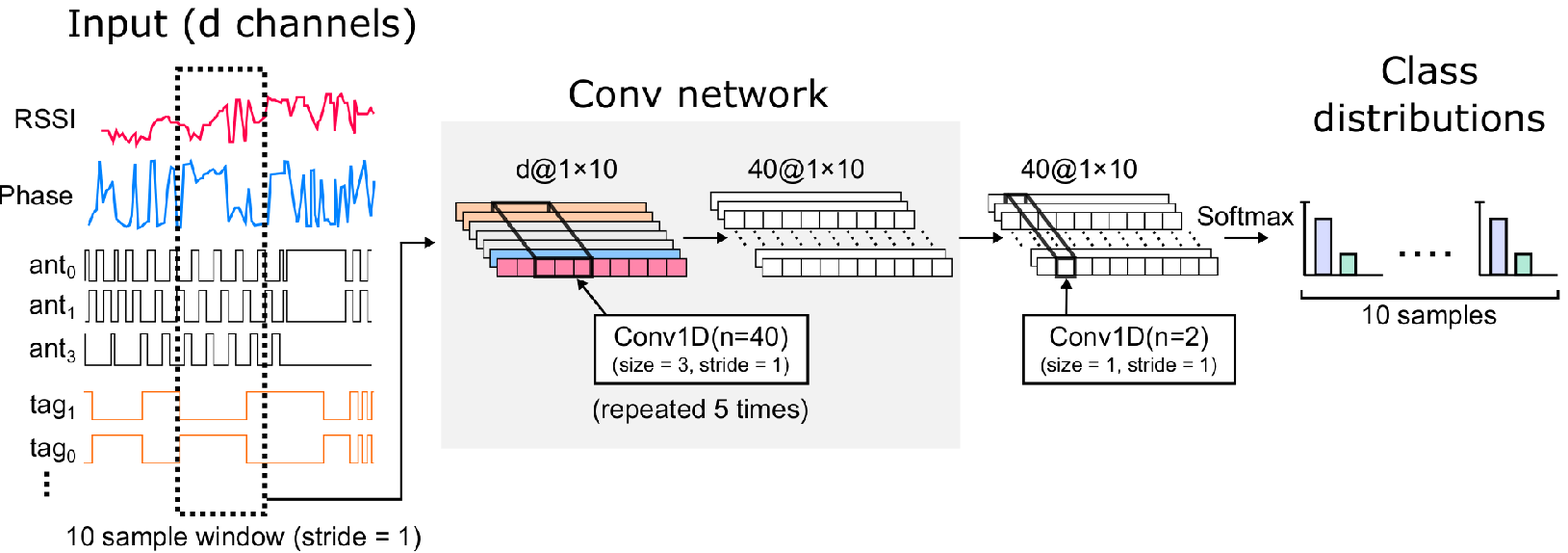}
		\label{fig:fcn_plot}
	}
	\caption{a) An overview of the ConvLSTM network used. A 20 sample window is fed into a network consisting of 3 1D-convolutional layers (each with a kernel size of 3, stride of 2, and 40 filters), followed by a dense RNN network (LSTM cells with 160 filters). A window stride of 1 is used to obtain a prediction for every input. We use ELU as the activation function and add dropout (0.5) at the final layer. This differs slightly from the original network described in~\cite{cnn-lstm-network}. We found these choices performed better with the low information content of the sensor data and the smaller dataset size. b) An overview of the FCN network we used. A 10 sample window is fed into a network consisting of 6 layers. The first 5 layers consist of 1D convolutions (filter size of 3, LeakyReLU activations) followed by a max pool, after these layers we add one dropout layer. We used a 1x1 Conv layer as the last layer. Similar to~\cite{yao2018efficient} we kept the stride of the Conv and Max-pool layers as 1 to ensure the output size matches the input size. Please see \cite{githublink2018} for further details.}

\end{figure*}
\vspace{1mm}
\noindent\textbf{Specific Features Engineered for the \sensor approach: }

We build the following features to extract the acceleration information available in a segment. The ID of the most recent tag reading ($a_i$) was considered as a feature as it approximates to a one-bit acceleration value. We included the relative ID count ($RI \in \mathbb{R}^2$), as a feature as it provides information  about the changes in the IDs or rate of change information---see Figure~\ref{fig:state1}---during the considered segment $S_i$.
In order to allow the classifiers to learn from RSSI patterns specific to each ID as shown in \fig{fig:state1}, we extracted RSSI features described previously with respect to each reported ID.
In total we extracted $15$ features for the \tagapproach approach and  $17$ features for \sensor approach.

\subsection{Neural Network approaches}

Instead of engineering features, current research suggests that we can learn features from kinematic sensors and build a classifier using deep neural networks using raw readings. We consider two state-of-the-art architectures: i) Deep Convolutional and LSTM Recurrent Neural Network; and ii) Fully Convolutional Neural Network architecture.
\vspace{1mm}

\noindent\textbf{Deep Convolutional and LSTM Recurrent Neural Network (ConvLSTM): } We design a network architecture following  \cite{cnn-lstm-network}. This network architecture has reported state-of-the-art performance on benchmark human activity datasets based on kinematic sensors. The convolutional layers learn to extract features under the independent and identically distributed assumption of the input segments while LSTM layers captures temporal dependencies in sequential data, such as the RFID data stream in our work. The network architecture we developed and used is described in \fig{fig:cnn_lstm_plot}.
\vspace{1mm}

\noindent\textbf{Fully Convolutional Neural Network (FCN):} We design a fully-convolution-network (FCN) following the design in~\cite{yao2018efficient}. This network architecture is more convenient to train than the LSTM based networks since it lacks any recurrent connections. In addition, the fully convolutional architecture has reported new state-of-the-art performance measures for benchmark human activity recognition datasets with body worn kinematic sensors. Further, the FCN architecture addresses issues with window label ambiguity---the multi-class window problem~\cite{yao2018efficient} where windows overlap multiple classes while ground truth labels are limited to selecting, for example, the majority class as the ground truth label for a given segment. The network architecture we designed for our classification problem based on~\cite{yao2018efficient} is described in \fig{fig:fcn_plot}.
\vspace{1mm}

As input, we use a minibatch size of 80, feeding the inputs [tag ID ($a$), antenna ID $aID$, $RSSI$, and phase ($\Psi$)] as individual channels to the network. We use a window size of 20 for the ConvLSTM network and a window size of 10 for the FCN. For the ConvLSTM based network, we unroll the network 40 steps. Both networks were trained using gradient descent with the Adam optimizer until convergence. The code and model parameters for the networks used will be available at \cite{githublink2018}.

\section{Statistical Analysis}
Our main objective is to reduce the missed and false bed-exit events. Therefore, we selected precision ($P$) and recall ($R$) for our evaluation and measure performance using the $\text{F}_1$-score ($F$) calculated as: $F={(2 \times P \times R)/(P+R)}$.

In order to evaluate bed-exit event recognition performance we defined a true positive (TP) bed-exit event as:  i) a bed-exit event that occurs no more than $\delta t$ time before the actual bed-exit; or ii) a bed-exit event that is recognized while the patient is actually \textit{out-of-bed}. False positive (FP) bed-exit events are incorrectly recognized bed-exit events based on the above definition of TP. Now, $P=TP/(TP+FP)$ and $R=TP/(TP+FN)$.

As our study participants were hospitalized older patients,  we observed during our trials that while sitting on bed, often, several attempts were required by patients to actually transition out of bed. Therefore, we analysed the sitting on bed durations ($D$) for patients before getting out of the bed and selected $\delta t = mean(D) + std(D) \approx 30$\,s. In the study, bed-exit alarms are recognized as TPs as long as the patient is \textit{out-of-bed} because people could fall during and after bed-exits~\cite{robinovitch2013video}; hence, knowing that the patient has left the bed will provide the opportunity for the nurses to intervene and possibly prevent a fall or provide immediate assistance in case of a fall.

We evaluate our performance measures using \textit{leave one patient out} cross validation. This validation approach is a participant independent testing scheme. This allows us to evaluate performance against a participant never seen during training or validation. Although this approach can show poor performance results, we can expect the results to be closer to that realized in a real world deployment.

\begin{table}[t]
\caption{Bed-exit Performance with Leave-one-out Cross Validation}
\centering

\resizebox{0.7\linewidth}{!}{
\begin{threeparttable}
	\begin{tabular}{|ll|l|l|l|}
	    \hline
		\multicolumn{2}{|c|}{\textbf{Approach}} &       \textbf{F-Score}       &       \textbf{Precision}       &      \textbf{Recall}      \\ \hline
		\textit{\tagapproach Approach} & LR & 0.68 & 0.53 & 0.96 \\
		 & FCN & 0.84 & 0.75 & 0.94 \\
		 & ConvLSTM & 0.87 & 0.93 & 0.83 \\ \hline
		\textit{\sensor Approach} & LR & 0.86 & 0.79 & 0.96  \\
		& FCN & \textbf{0.90} & 0.85 & 0.97 \\
	    & ConvLSTM & 0.87 & 0.98 & 0.79  \\ \hline
	\end{tabular}
\end{threeparttable}
}

\label{tbl:performance}
\end{table}

\begin{table}[t]
\caption{Bed-exit Recognition Performance for Each Patient Using FCN}
\centering
    \begin{tabular}{|c|c|c|c|c||c|c|c|}
    \hline
    & & \multicolumn{3}{c||}{\textbf{\sensor}} & \multicolumn{3}{c|}{\textbf{\tagapproach}} \\ \hline
Patient ID	& Actual	& TP	& FP	& FN	& TP	& FN	& FP \\ \hline
0	& 3	        & 2	& 0	& 1	        & 2	& 0	& 1 \\ \hline
1	& 3	        & 3	& 0	& 0	        & 3	& 0	& 0 \\ \hline
2	& 3	        & 3	& 0	& 0	        & 3	& 0	& 0 \\ \hline
3	& 2	        & 2	& 0	& 0	        & 2	& 2	& 0 \\ \hline
4	& 3	        & 3	& 0	& 0	        & 3	& 3	& 0 \\ \hline
5	& 1	        & 1	& 1	& 0	        & 1	& 0	& 0 \\ \hline
6	& 2	        & 2	& 1	& 0	        & 2	& 1	& 0 \\ \hline
7	& 6	        & 6	& 0	& 0	        & 6	& 1	& 0 \\ \hline
8	& 3	        & 2	& 0	& 1	        & 3	& 0	& 0 \\ \hline
9	& 3	        & 3	& 4	& 0	        & 1	& 3	& 2 \\ \hline
10	& 3	        & 3	& 0	& 0	        & 3	& 0	& 0 \\ \hline
11	& 2	        & 2	& 0	& 0	        & 2	& 0	& 0 \\ \hline
12	& 3	        & 3	& 0	& 0	        & 3	& 0	& 0 \\ \hline
13	& 3	        & 3	& 0	& 0	        & 3	& 0	& 0 \\ \hline
14	& 3	        & 3	& 0	& 0	        & 3	& 3	& 0 \\ \hline
15	& 3	        & 3	& 0	& 0	        & 3	& 0	& 0 \\ \hline
16	& 3	        & 3	& 0	& 0	        & 3	& 2	& 0 \\ \hline
17	& 3	        & 2	& 0	& 1	        & 2	& 0	& 1 \\ \hline
18	& 3	        & 3	& 0	& 0	        & 3	& 1	& 0 \\ \hline
19	& 3	        & 3	& 4	& 0	        & 3	& 1	& 0 \\ \hline
20	& 3	        & 3	& 0	& 0	        & 3	& 0	& 0 \\ \hline
21	& 4	        & 4	& 0	& 0	        & 4	& 3	& 0 \\ \hline
22	& 5	        & 5	& 2	& 0	        & 5	& 1	& 0 \\ \hline
Total & 70	    & 67 & 12 & 3       & 66 & 21 & 4 \\ \hline
    \end{tabular}
    \label{tbl:perpatient}
\end{table}

\section{Results}

Table~\ref{tbl:performance} illustrates performance obtained for bed-exit event recognition (defined as in-bed to out-of-bed transition, see~\cite{githublink2018} for the alarming algorithm) using leave one patient out cross validation.

When considering all three classification methods, according to the mean F-Score, \sensor approach has performed better than the \tagapproach approach. We can expect \sensor approach to perform better as a consequence of the information from the low resolution acceleration data embedded in ID modulations of the \sensor approach (see \fig{fig:state1}).

We further analyzed the bed-exit event recognition performance in terms of TPs and FPs for each patient. The results are shown in Table \ref{tbl:perpatient} for the best performing method, FCN. We can see a significant reduction in FPs with the \sensor approach. This is due to the additional information extracted from the low resolution acceleration data in the \sensor approach, as shown during an activity transition in \fig{fig:state1}, as well as the RSSI pattern associations with the activated ID during static postures such as lying and posture transitions.

Upon close inspection of the instances where false bed-exit events or missed bed-exit events are reported, we identified that they are predominantly due to inadequate tag readings while lying in bed and during getting out of bed. The missed tag readings are a consequence of the \sensor being attached to a single shoulder of a patient's loosely fitted garment and not being adequately exposed to the antennas. This can occur as the patient is lying on their side on the bed with the shoulder on the mattress. The lack of data or the sparseness in data consequently led to false or missed bed-exit events.

\begin{figure}[t]
	\centering
	\includegraphics[width=0.80\linewidth]{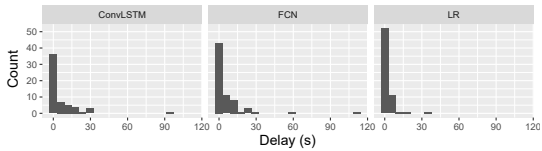}
	\caption{Bed-exit event recognition delays for the \sensor approach}
	\label{fig:delay}
\end{figure}

In addition to predicting bed-exits correctly, the system should ideally report bed-exits in a timely manner---with low latency. Figure~\ref{fig:delay} illustrates the distribution of delays with respect to correctly identified bed-exit events (i.e. TPs). According to the distribution of delays we observed that, for all approaches, the majority of the bed-exit events (more than 80\% for FCN, 65\% for ConvLSTM, and 91\% for LR) are identified within a period of 10\,s of an actual event. Notably, a classifier that generates a large number of false positives will likely result in less delays due to constant alarm predictions. This can be seen with the results of LR where nearly 90\% of alarms are with in 3~s. However, we can see from Table~\ref{tbl:performance} that LR for the \sensor approach has high recall but lower precision values indicative of a larger number of false alarms (i.e. FPs).

\begin{figure}[t]
	\centering
		\includegraphics[width=0.36\linewidth]{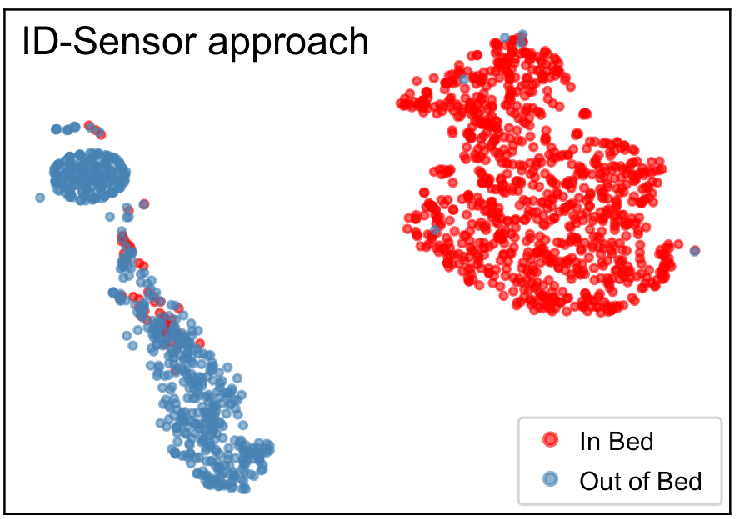}
		\includegraphics[width=0.36\linewidth]{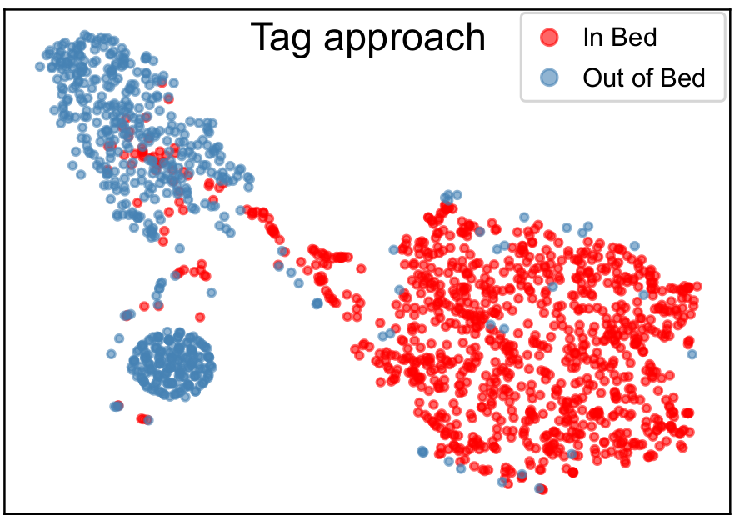}
	\caption{A 2D visualisation of the learned feature spaces using t-distributed stochastic neighbour embedding (t-SNE) with FCN.}
	\label{fig:tsne}
\end{figure}

We  visualise  the  high-dimensional feature space to understand the usefulness of ID sensor data in Fig.~\ref{fig:tsne}. Here, the segment embeddings for each activity are clustered together with different activities clearly separated in the feature space for the \sensor approach compared with the \textit{Tag approach}.

\section{Related Work}\label{sec:relatedwork}

Based on the sensor deployment strategy,  we categorize the  existing  bed-exit recognition systems  broadly into: i)~environmental sensor-based approaches or device-free methods; and ii)~body-worn sensor-based approaches. Given the clinical context of our study, we focus mainly on approaches relevant for detecting bed egress motions as opposed to general activity recognition or human motion detection.

\subsection{Environmental sensors}
Studies  have investigated the performance of bed-exit alarming systems based on one or multiple sensors strategically placed on or around
the bed \cite{bruyneel_detection_2011, hilbe_development_2010, vass_refine_2009, capezuti_bed-exit_2009}.
Most of these methods involved pressure sensors or pressure mats. In general, pressure mats need daily maintenance to check correct functionality as they are subject to constant mechanical stress and are highly likely to move from their ideal placement on the bed. Moreover, pressure mats require disinfection because of possible exposure to body fluids, and protocols for controlling infections. Additionally, pressure sensors were shown to be unreliable for patients lighter than 45.4~kg (100 lbs)~\cite{capezuti_bed-exit_2009}. In fact, two recent randomised control trials evaluating pressure sensor alarms systems to prevent falls found no significant reduction in falls~\cite{shorr_effects_2012,sahota2013refine}.
Both studies reported high `false alarms' (incorrectly identified alarms) as a significant reason for the negative results.

In contrast Hilbe et al. \cite{hilbe_development_2010} used bed rails fitted with pressure sensors for bed-exit recognition. However, bed rails can potentially increase harm where a fall may then occur from a greater height resulting in more serious injury \cite{healey_bedrails_2009} and are no longer considered best practice.

Bruyneel et al. \cite{bruyneel_detection_2011} proposed the use of multiple types of commercially available sensors, placed under the bed. These included three sensors to measure the temperature,  two piezzo-electric sensors to measure the body movements and three resistive sensors to measure the presence or absence of the patient. The system was evaluated with young healthy participants. Notably, the temperature sensor resulted in a response delay of up to two minutes while it required more than one hour for the system to reach equilibrium between body and mat temperature.

Additionally, commercially available pressure sensor arrays attached to bed mattresses~ \cite{arcelus2011measurements} or RFID tags mounted onto walls~\cite{yao2018compressive} have  been used  to analyse   sit-to-stand motions.  Information from pressure sensor arrays were viewed  as image data to identify different phases of the sit-to-stand transitions.  Although the focus of these studies was not bed-exit recognition, a sit-to-stand transition may be considered to represent a bed-exit at times~\cite{wickramasinghe2017sequence}.  However, further evaluations  are required to measure performance for bed-exit recognition, especially with frail older people. The studies above attached sensors to infrastructure rather than individuals, limiting their detection to a specific activity or physical area whilst not solving the \textit{person distinguishability problem} of device free methods~\cite{wang2018rf}.

\subsection{Body-worn Sensors}

Research studies have looked at kinematic sensors \cite{najafi2003ambulatory, godfrey_activity_2011,  wolf_development_2013,Wang2013,wickramasinghe2017sequence,roberto-pone.0185670} for  ambulatory monitoring because of the added possibility of monitoring a person in multiple locations. Most of these studies follow the pioneering study in \cite{najafi2003ambulatory} where  a kinematic sensor, composed of one miniature
piezoelectric gyroscope and two miniature accelerometers, was attached to a person's sternum to monitor activities such as walking and sit-to-stand posture transitions.

However, there is limited research work focused on movement sensor alarm systems for bed-exit recognition \cite{wolf_development_2013,roberto-pone.0185670,visvanathan2017effectiveness}. Researchers in~\cite{wolf_development_2013} used a  battery powered acceleration sensor that was strapped with a bandage to the thigh to acquire motion information  to identify bed-exit events. The clinical trial in~\cite{visvanathan2017effectiveness} proposed the use of a wearable Bluetooth sensor inserted into a patient vest for determining bed and chair egress movements to realize an alerting system.

Even though battery powered body worn devices generally provide rich sensor data, they are expensive, bulky, obtrusive and require maintenance such as changing or replacing batteries~\cite{visvanathan2017effectiveness}. Further, evidence in the literature highlights the preference for small, unobtrusive and easy to operate monitoring devices by older people\cite{steele2009elderly}.

Recently, wearable and batteryless Computational RFID (CRFID) tags capable of supporting embedded sensors have been used to recognize bed and chair egress movements~\cite{wickramasinghe2017sequence,roberto-pone.0185670}. These studies mainly rely on a MEMS 3D accelerometer sensor data to recognize bed and chair egress movements.
Although we have seen the commercialization of CRFID technology recently, a wearable CRFID is still undergoing research and development. Further, the per-unit cost of CRFID devices are still several tens of USD. Moreover, these studies show that wirelessly powering a MEMS 3D accelerometer remains a challenging task and leads to loss of information and highly sparse data streams that affect performance of activity recognition algorithms~\cite{roberto-pone.0185670}.

\section{Discussion and Conclusion}\label{sec:discuss-conclude}	We have designed a sensing approach using mature COTS UHF RFID technology without the need for a high resolution kinematic sensor employed in previous studies. Our approach provides significant advantages when compared to CRFID devices, battery powered body worn sensors and pressure sensors.
COTS passive RFID tags are: \textbf{\textit{i)}}~small in size, thus increasing  the possibility for integrating the tags into hospital gowns; \textbf{\textit{ii)}}~low in cost---0.07 to 0.15 USD~\cite{passiveRFIDJournal} and batteryless, thus providing greater economic advantages and increasing the possibility  to dispose the tags to support hygiene protocols in hospitals; and \textbf{\textit{iii)}}~able to solve the problem of distinguishing individuals and monitoring individuals in multiple locations.

We investigated the efficacy of our \sensor approach for recognizing hospitalized fail older people's motions to alert on bed-exits. We have developed a fully convolutional neural network architecture capable of learning information from RFID data streams as opposed to previous applications in wearable sensor data from kinematic sensors. The \sensor approach with the FCN dense labelling and prediction approach depicted the highest performance (F-score of 86\%).  Further, our data was collected in either single- or double-bed rooms, thus suggesting that the approach is \textit{agnostic to the environment across similar antenna deployments}.
\vspace{1mm}

\noindent\textbf{Comparisons:} In the recent past, several bed-exit systems have been developed. Table\,\ref{tbl:previous} summarizes the results of previous bed-egress movement recognition approaches, together with our study results. We have excluded pressure mats given the lack of clinical evidence for their efficacy. It is difficult  to make a fair comparison with these due to: \textbf{\textit{i)}}~differences in the experimental setting such as the characteristics of the study participants and the duration of the study; and \textbf{\textit{ii)}} differences in the performance evaluation measures used. Nevertheless, we can see that the \sensor performs comparably better than methods tested with older people.

\begin{table}[t]
\caption{Performance of Previous Bed-exit Movement Alarm Methods}
\label{tbl:previous}
\centering
\resizebox{1\linewidth}{!}{
\begin{threeparttable}
\begin{tabular}{llll}
	\toprule
	Bed-exit recognition approach                     & Precision & Recall  & Participants' age (years)     \\ \midrule
	Hilbe~\etal~\cite{hilbe_development_2010}           &           & 96\%        & 18-60                  \\
	Bruyneel~\etal~\cite{bruyneel_detection_2011}       & 100\%     & 91\%        & 37$\pm$9 and 45$\pm$11 \\
	Najafi~\etal~\cite{najafi_ambulatory_2003}\tnote{*} &           & 93\%          & 66$\pm$14          \\
	Godfrey~\etal~\cite{godfrey_activity_2011}\tnote{*} &           & 83\%          & 77.2$\pm$4.3          \\
	Torres~\etal~\cite{roberto-pone.0185670}   & 66.83\%     &  81.44\% & 71 to 93 \\
	\textbf{Ours}                                  &  85\%     &  96\% & 84.4 $\pm$5.3 \\ \bottomrule
\end{tabular}
\begin{tablenotes}
\item[*]a sit-to-stand posture transition was considered as a bed-egress.
\end{tablenotes}
\end{threeparttable}
}
\vspace{-10pt}
\end{table}

Studies~\cite{hilbe_development_2010,bruyneel_detection_2011} considered bed exits and reported recall values of over 90\%. However, performance reported was based on experiments with young and middle-aged adults while absence of precision results means that we are unable to comment on false alarms as a function of all alarms. Notably, the empirical methods used were developed and tested with the same dataset; consequently, yielding optimal heuristic measures for the particular dataset. In contrast to a low-cost batteryless RFID tag that can be disposed if required and does not require maintenance re-charging batteries, the sensor required disinfection or thorough cleaning because of possible exposure to body fluids or infection control.

Our results are higher in comparison to studies with body worn batteryless CRFID devices with a high quality 3D accelerometer sensor evaluated with hospitalized older people~\cite{roberto-pone.0185670}. Whilst the high resolution accelerometer provided more information, the RF powering issues from various postures from older people and movement such as rolling out of bed by patients resulted in highly sparse data streams. As a consequence, the results show a relatively larger number of false positives compared to our study (66.85\% vs. ours 86\%).
\vspace{1mm}

\noindent\textbf{Limitations: } Although our results are promising, our study is not without limitations. During the pilot study, experiments were conducted for few hours using broadly scripted activity routines. Therefore, it is important to validate the results using longer duration trials (such as days opposed to hours), at night as well as daytime. Additionally, we have not evaluated the mean time before failure of the mechanical switch in the \sensor.
While the \sensor is low cost and has the potential for disposability and textile integration, deploying RFID reader and antenna infrastructure across a hospital is costly and was not discussed here. 	However, RFID infrastructure (which consists of RFID antennas and readers) is increasingly being deployed in hospitals for patient and asset tracking  \cite{prater2017enhancing}. Hence, such existing infrastructure is envisioned to be also utilized for patient activity monitoring using an \sensor.
	\vspace{1mm}

\noindent\textbf{Future Work: } We leave for future work: \textbf{\textit{i)}} performance evaluation with a sensor integrated hospital gown; and \textit{\textbf{ii)}} the use of two \sensor tags over both right and left shoulders of a patient. As a result of the additional information, we expect  the tag reading rate to improve and consequently the number of missed tag readings to reduce further to yield higher bed-exit event recognition performance. Further, to establish efficacy,  it is  necessary to evaluate our approach using a larger study such as a randomised controlled trial.

\section*{Acknowledgments}
Our study was supported by a grant from the Hospital Research Foundation and the Australian Research Council (DP160103039). We would like to thank Dr. S. Nair and Mr. S. Hoskin for supporting the trial at the Queen Elizabeth Hospital, South Australia. We would like to acknowledge Mr. R.L. Shinmoto-Torres for collecting and labelling the dataset.

\bibliographystyle{IEEEtran}
\bibliography{IEEEabrv,main}

\end{document}